# 10 Inventions on key guides and Keyboard templates
## -A study based on US patents

**Umakant Mishra**

Bangalore, India
umakant@trizsite.tk
http://umakant.trizsite.tk

**Contents**

# 1. Introduction

Keyboard is the most important input device for a computer. A keyboard has several keys including character keys, numeric keys, navigation keys, function keys and other special keys. As there are so many function keys and each function key can have multiple functions when used with control, shift and alt keys, it is difficult for a user to remember the functionality of the function keys. We need a mechanism to indicate the operations assigned to each function key for different software programs. A keyboard guide or template is used for this purpose.

**1.1 Need for using a keyboard template**

**Conventional keyboards**

A conventional keyboard generally contains a set of function keys which are supposed to do different activities in different software. Some of these special keys are:

- Function keys generally found above the character keys.

- Function keys used in combination with shift, ctrl and alt keys.

- Character keys in combination with ctrl, alt and other special keys, such as, ^c and ^v are used for copy and paste in some software.

- Page up, page down, insert, delete etc. used in combination with other keys.

Each of these special keys is used for specific operations under different software environment. This leads to hundreds of functionalities for these keys which is not possible to remember on the part of any common user. It is necessary for him to refer to a template.

**Reduced key keyboards**

There are some special keyboards which contain less number of keys, such as, 4 or 8 keys to generate all the characters and functions available in a full sized keyboard. In such cases, each key generates different characters in different modes. This mechanism needs a template to refer even for basic typing of characters.

The hand held computers and pocket computers have very less space for providing adequate keys. In many cases there are special keys which play different roles when pressed with different characters. A template is necessary to remind the user about these special roles.

**Other situations**

- New users and children may need to refer to key guides and templates.
- A special keyboard having additional keys for specific functions.
- Software that requires function keys to be used more often.

In all these situations, it is necessary to provide templates to help the user to use these special keys to get benefit of those.

### 1.2 Various options of keyboard guide and keyboard templates

As we saw in the situations above, a user may prefer to use a keyboard guide in various circumstances. However, a key guide mechanism can be different in different situations. Some of the popular key guide mechanisms are as follows.

- A reference manual- that is not very convenient as the user has to open the pages of the book.

- Putting stickers on the keys- this method is ok as long as the user uses only one software. As the role of the keys varies from software to software, the meanings of the stickers does not hold good for other software.

- A template bar above the function keys- is good but there no space for sticking multiple bars for different software.

- A dynamic template bar above the function keys which contains various templates for various software. The user can change templates depending on the software used.

- An electronic template displayed on the keyboard. this is good but, the keyboard should have a display system.

- A software template displayed on the screen. This method was very popular during nineties. The disadvantage is that it occupies valuable screen space.

### 1.3 Analysis of keyboard templates from TRIZ perspective

Particularly in hand held computers, there is a very limited room available for accommodating all the keys used in a large keyboard. Hence it is required to reduce the number of keys without sacrificing the performance of the keyboard operation. Reducing number of keys on a keyboard often requires pressing of two or more keys to generate a desired command which creates confusion with the user and sometimes leads to error **(Contradiction)**.

This problem is addressed by integrating templates (or operating instructions) along with the keyboard **(Principle-8: Counterweight)**.

It is better to have a dynamic template which can be changed according to change of the software in use **(Principle-15: Dynamize)**.

It is still better to display the templates automatically as and when the software environment is changed as in case of a software template **(Principle-15: Dynamize, Principle-25: Self service)**.

It will be even better to display a template without occupying the valuable screen space by using various methods.

## 2. Inventions on key guides and keyboard templates

### 2.1 Hand held computers with alpha keystroke (5067103)

**Background problem**

It is known to produce more entries from fewer keys by changing modes. Although the number of keys can be reduced, the changing of modes from time to time is difficult and likely to cause errors in operation. On a hand held computer there is little room for visually accessible operating instructions for so many choices. It is necessary to provide improve instructions for entry of many different input commands from a limited number of keys.

**Solution provided by patent**

Lapeyre disclosed a method of displaying instructions for keyboard operations (assigned to The Laitram Corporation, issued November 1991). The instruction is meant for keyboards having very few keys to operate large number of characters by using two or more strokes per character. The invention displays multi-choice keystroke instruction menus dynamically when the user operates the keyboard.

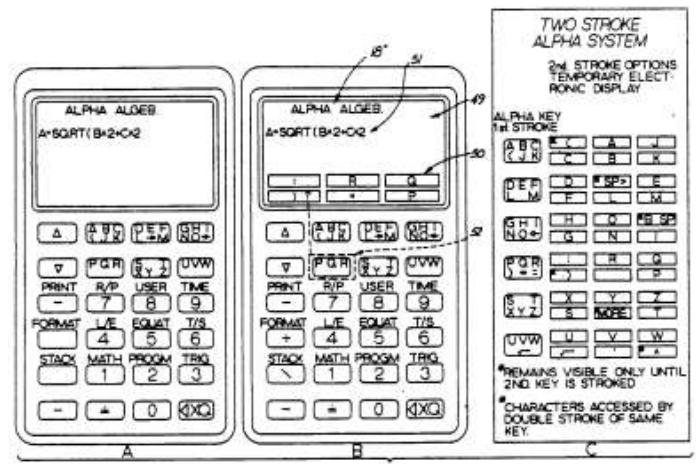

When the instructions are displayed, the display screen is divided into two separate regions for displaying keyboard instructions and data. According to the invention, the instructions are displayed only when necessary and not always.

**TRIZ based analysis**

The main objective of the invention is to provide improved instructions for keyboard operation, to eliminate the ambiguity of the keyboard keys in different modes **(Principle-8: Counterweight).**

The multi-choice keystroke instruction menus are dynamically displayed when the user operates the keyboard **(Principle-15: Dynamize)**.

It divides the screen and creates a temporary display area to display the keyboard instructions **(Principle-1: Segmentation)**.

The instructions are displayed on the screen only when required and not always **(Principle-15: Dynamize)**.

### 2.2 Computer keyboard function key guide (5080516)

**Background problem**

The modern keyboards have several control keys and function keys for special operations. Many software programs provided by different venders use these function keys for various different purposes. In some cases the function key guides relating to the particular software are provided in a strip, which are placed near the function keys for easy reference. These key guides provide directions to the user indicating the particular function within the software. But this method does not solve the problem for users using more than one software program. If the user interchanges the key guides for every change of program, it becomes difficult for the user and causes damage to the key guides.

**Solution provide by the patent**

Ward disclosed a key guide assembly to identifying functions of control function keys on a computer keyboard (Patent 5080516, assignee- Sarasota Technologies Inc., Issued Jan 1992). One side of the assembly can hold multiple key guides, the other side is meant to attach to the keyboard. This solution is inexpensive and adaptable to any type of keyboard.

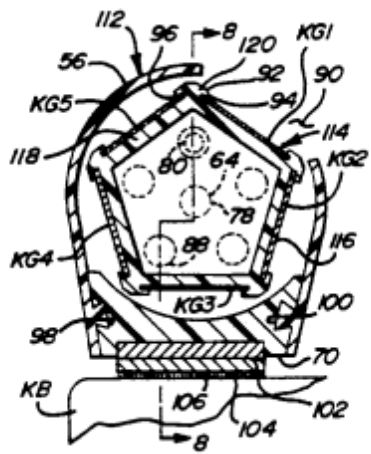

The software key guide include templates made of thin material and having a plurality of columns associated with each function key and a plurality of columns providing individual boxes for marking the function thereon.

### TRIZ based analysis

The objective of the invention is to assist users in keyboard operation **(Principle-8: Counterweight)**.
The solution is inexpensive **(Principle-27: Cheap and disposable)**.
The function key templates are made of thin materials **(Principle-30: Thin and Flexible)**.
The invention suggests a coloring method; the rows are of uniform color and different from an adjacent row so that each box associated with a particular function key is of a different color. **(Principle-35: Color change)**.

### 2.3 Programmable hand held computers operable with two-strokes per-entry alpha with instruction menus on temporary viewing screen (5124940)

### Background problem

A computer keyboard having a few keys controlled by a multimode processing system needs operational instructions. In handheld computers there is very less space for having enough keys, and obviously no room for displaying instructions to operate the keyboard.

### Solution provided by the invention

Lapeyre disclosed (patent 5124940, assignee-The Laitram Corporation, issued June 1992) a computer organized temporary display panel that displays keystroke selection indicia for the different keyboard operating roles. The display panel is programmed by the computer to show the instruction guide of the keys in the current operating role of the keys.

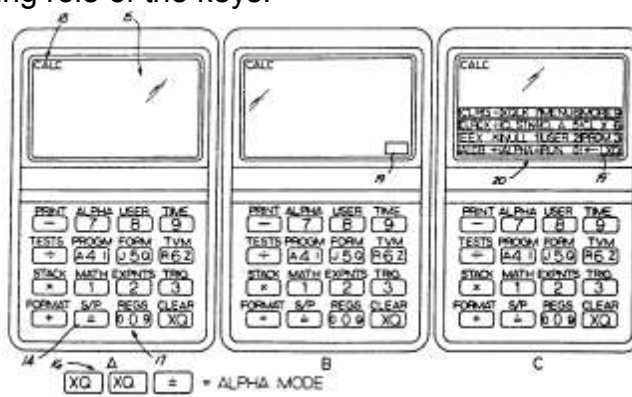

This permits detailed keystroke instructions to be available for performing hundreds of operations for which the computer is capable.

**TRIZ based analysis**

As each character in a small keyboard requires multiple keystrokes the invention provides operational guideline to remove the ambiguities of sequence of keystrokes **(Principle-8: Counterweight)**.

Multi-choice keystroke instruction menus are dynamically displayed as the modes and stroke sequencing of the computer take place **(Principle-15: Dynamize)**.

The display screen is divided into separate regions for instruction indicia and data **(Principle-1: Segmentation)**.

The instruction menu appears only when a key is pressed and disappears when the character is complete for the next keystroke **(Principle-34: Discard and recover)**.

## 2.4 Stacked computer keyboard function key multiple template retainers (5144303)

**Background problem**

The same function keys are used for different purposes in different operating systems or different applications. For example, the function key F4 may have one meaning in Microsoft Word and a different meaning in a Lotus Spreadsheet. The user sometimes presses wrong function keys because of his habit with another program. It is difficult for the user to remember the meaning of the function keys. There is a need for a mechanism to remind the user about the meaning of the function keys.
Although, we can display the function key labels at some place on the screen, that will occupy valuable screen space.

**Solution provided by the invention**

The solution is a template retainer invented by Ronald Purcell (Patent 5144303, Issued in Sep 92). The template retainer comprises flexible strips constructed of polyester film. The strips are relatively flexible and extremely durable. The sides of the trips are connected with the keyboard with adhesive in such a manner that the strips may be swung in order to be selected. These strips contain the function key templates or identity strips for swinging movement between supported positions near the function keys. The user changes the strips selectively to move from one temple to other.

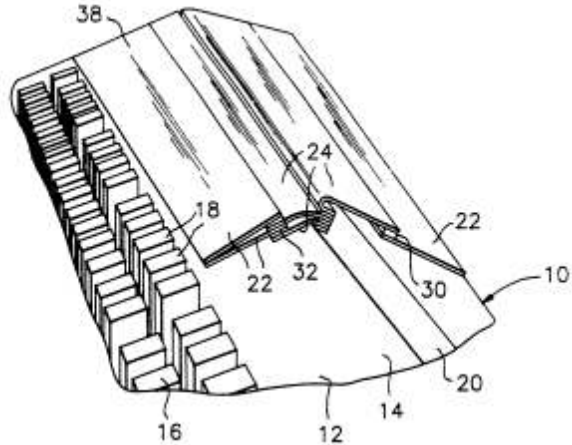

**TRIZ based analysis**

Ideally the function keys themselves should inform about their function in different applications **(Ideal Final Result)**.

The invention provides a mechanism to change the function key templates as required for different applications **(Principle-15: Dynamize)**.

The invention uses a stacking arrangement of keeping one template over the other **(Principle-7: Nested Loop)**.

**2.5 Stand for displaying computer keyboard function key guides (5144763)**

**Background problem**

The conventional keyboards have a series of function keys that are used by different software for doing special operations. The type of function performed by any particular function key varies from software to software. While switching from one software to another, it is difficult for the operator to remember the changes that have taken place in the function keys.

In order to minimize some of the confusion concerning the function keys, some companies have devised labels to be affixed on the upper surface of each function key. But this method does not allow changing the labels when the user works on different software.

**Solution provided by invention**

Calhoun disclosed a stand for displaying function key guides (Patent 5144763, assignee-Nil, Issued Sep 1992). The stand includes a tubular, open-ended sheath preferably having three rectangular shaped, transparent sides. The printed key guides can be easily inserted into the sheath, which is visible to the keyboard operator. The additional key guides may be stored within the interior of the sheath.

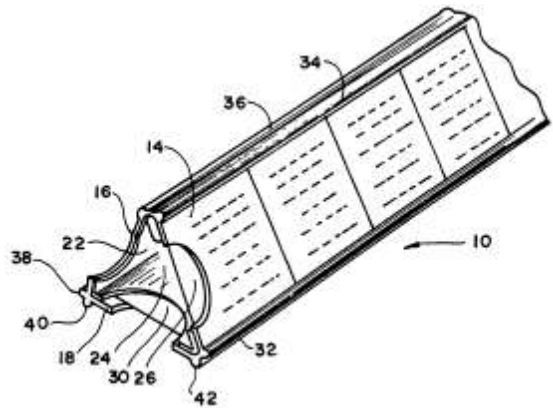

**TRIZ based solution**

The stand for displaying the key guides intends to reduce users' confusion on function keys **(Principle-8: Counterweight)**.

The stand has three sides to hold three guides **(Principle-17: Another dimension)**.

The guides may be selectively inserted into or removed from the interior of the sheath of the stand **(Principle-34: Discard and recover)**.

**2.6 Electronic keyboard template (5181029)**

**Background problem**

As there are so many function keys having different functions when used in different software and when used with control, shift and alt keys, it is difficult for a user to remember the functionality of the function keys.

There are some inventions on using plastic sheet templates for function keys. But this solution is limited as each new application can have a different template. Even if the users are given new sheets of templates for each new application, they are not flexible to reflect the custom definitions of the function keys made by the user.

**Solution provided by the invention**

Jason Kim invented an electronic keyboard template (Patent 5181029, assigned to AST Research, Issued in Jan 93) for use with software application programs. The electronic template includes an LCD display screen for displaying icons representing the operations performed by the function keys. The invention includes a microprocessor, memory and a program to control and display the icons for the template.

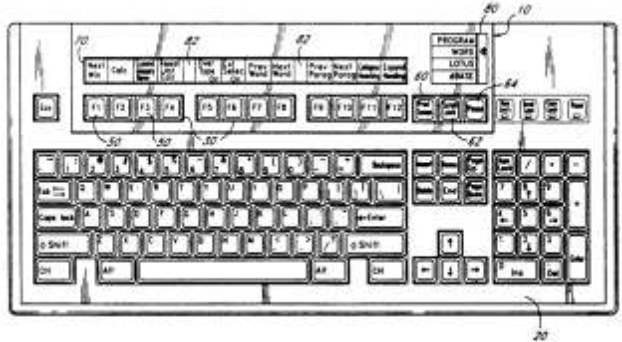

This electronic keyboard template overcomes the limitations of prior manual keyboard templates. As the templates are displayed by the control circuitry, it allows to be configured for a new application or changes made for a user-defined customization.

**TRIZ based analysis**

The invention displays function keys definitions to the user for reference during keyboard operation **(Principle-8: Counterweight)**.

The templates are changed according to the change of application **(Principle-15: Dynamize)**.

The templates are changed automatically when the meaning of the function keys are changed **(Principle-25: Self Service)**.

### 2.7 Computer keyboard and template holder (5497970)

**Background problem**

Although there are some inventions on the keyboard holder there is no invention on holding a keyboard template along with the keyboard. There is a need for an improved keyboard and template holder.

**Solution provided by the invention**

William Pursell disclosed a keyboard and template holder (Patent 5497970, Mar 1996). The invention attaches the template holder and the paper holder along with the keyboard holder. The template holder uses a plurality of plastic rectangular sleeves coupled to the binder rings, each sleeve having space for inserting a template.

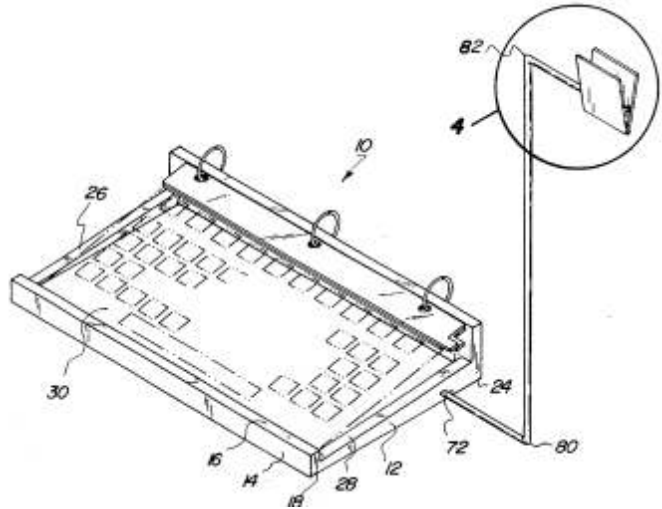

**TRIZ based analysis**

The invention discloses templates hanging on a ring so that the user can easily change them by rotating the templates **(Principle-15: Dynamize)**.

**2.8 System for reconfiguring a keyboard configuration in response to an event status information related to a computer's location determined by using triangulation technique (5867729)**

**Background problem**

Conventionally keyboards have dedicated function keys and various Alt or Ctrl key combinations, which invoke special functions or actions without needing the user to type a long command for the action. The function keys and special keys are differently configurable for different programs. Windows Program Manager provides a shortcut key facility to launch specific applications using function keys.

However this facility suffers from the drawback that the shortcut key definitions are valid only when Program Manager is active. The user may confuse to press some key combinations when the definitions are changed by another application is active.

**Solution provided by the invention**

Glenn Swonk discloses the solution to this problem (US patent 5867729, assigned to Toshiba America Information Systems, Feb 99). According to the invention, the definition of function keys and special keys are stored in a lookup table. A user interface detects the keyboard input and verifies its definition in the lookup table. If it has one single definition then it launches the application. If it has multiple definitions then the user is presented with a pop-up list to select an application for launching.

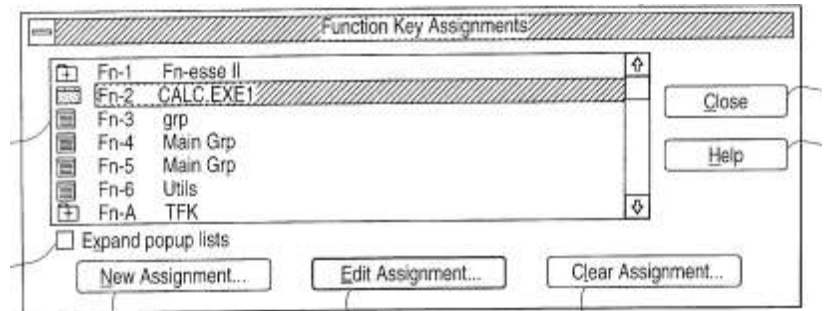

This saves user from making mistakes by pressing wrong key combinations in a changed environment.

**TRIZ based analysis**

The same function key should be definable for multiple functions so that the user can use it for various purposes. But defining the same function key for different functions will create confusion in the user. **(Contradiction).**

The invention displays a list of functions to choose when the same function key or special key is assigned to different functions in different applications **(Principle-9: Prior Counteraction)**.

**2.9 Computer keyboard system enabling users to locate keys of letters, radicals and phonetic symbols quickly (5954437)**

**Background problem**

The keys on a traditional keyboard are so scattered that it is difficult for the new users to quickly locate the keys on the keyboard. This drawback of the layout creates computer-phobia with the first time users. It is necessary to provide a user-friendly keyboard system.

**Solution provided by the invention**

Wen-Hung invented an improved keyboard (Patent 5954437, Sep 99) which uses different colors to improve identification of keys.

According to the invention, the colors of keys on the keyboard are arranged in groups according to five different patterns. Different sets of keys shares different colors, e.g., first set (A-G) shares one color, the second set (H-N) shares another color, the third set (O-R) shares another color and so on. Besides that, the left hand area and right hand area of the keyboard is divided by a demarcation line by using a different color for Y, H and N.

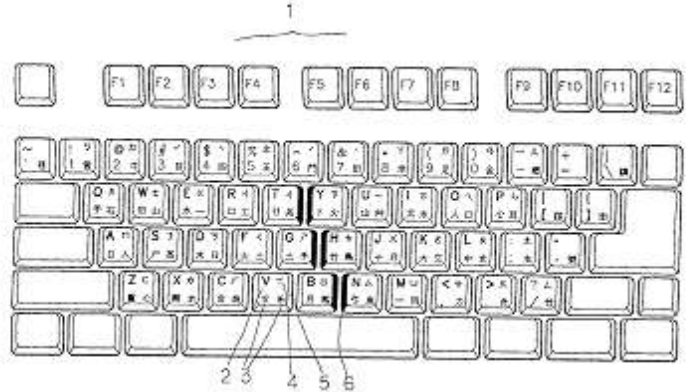

This coloring scheme helps user to locate keys on the keyboard.

**TRIZ based analysis**

The keys should help locating themselves **(Ideal Final Result)**.

The invention uses different colors for different sets of keys for easy identification **(Principle-32: Color change)**.

**2.10 Computer keyboard enhancement kit (6382854)**

**Background problem**

The dispersed structure of the keys on a keyboard makes it difficult for a beginner or a school child to find particular characters on the keyboard. Besides, only capital letter labels confuses a child to find the small letters in the keyboard. There is a need to enhance the keyboard for making the keys easily searchable by small children.

**Solution provided by the invention**

Morelos disclosed a method of detachable key replacements (Patent 6382854, May 2002). The key replacements are attached to the top surface of a transparent plastic jacket that is positioned on top of their corresponding letter keys on the computer keyboard.

The detachable keys will have more shapes and colors than the letter keys of a standard keyboard to make it easy for small children. The shapes and colors follow a specific pattern scheme wherein the letters of the same hand and row keys have key replacements with the same color but with varied shapes.

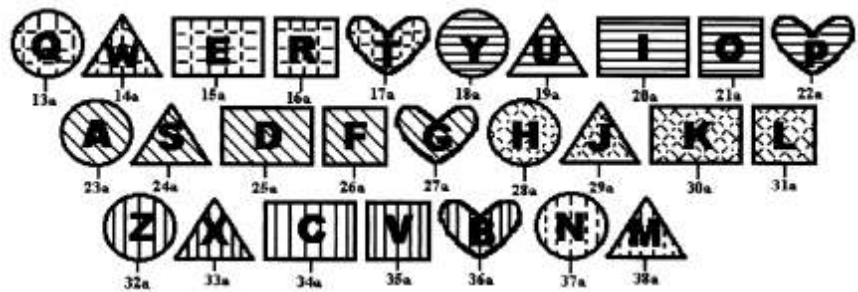

**TRIZ based analysis**

The invention makes use of colors to make the keys prominent **(Principle-32: Color Change)**.

Use replaceable key tops instead of building special keyboards for children **(Principle-34: Discard and Recover)**.

## 3. Summary and conclusion

The key guides and keyboard templates are useful to help user in using function keys, function keys in combination with shift, control or alt keys, other special keys and key combinations.

The method of displaying key guides can be manual (the user changes the key guide) or can be automatic (the key guide is automatically when the key definitions change), it can be paper based (to be pasted or hanged near the keyboard) or can be displayed on the screen.

Some key guide mechanisms use colors and beeps to differentiate between different types of keys.

## 4. Reference:

1.  US Patent 5067103, "Hand held computers with alpha keystroke", invented by Lapeyre, assignee- The Laitram Corporation, issued Nov 1991.

2.  US Patent 5080516, "Computer keyboard function key guide", invented by Ward, assignee Sarasota Technologies Inc, issued Jan 1992.

3.  US Patent 5124940, "Programmable hand held computers operable with two-strokes perentry alpha with instruction menus on temporary viewing screen", invented by Lapeyre, Assigned to The Laitram Corporation, Issued in June 1992.

4.  US Patent 5144303, "Stacked computer keyboard function key multiple template retainers", Ronald Purcell, assignee-Nil, issued Sep 92.

5.  US Patent 5144763, "Stand for displaying computer keyboard function key guides", invented by Calhoun, assignee-Nil, issued Sep 1992.

6.  US Patent 5181029, "Electronic keyboard template", Jason Kim, assigned to AST Research, Jan 93

7.  US Patent 5497970, "Computer keyboard and template holder", William Pursell, Mar 1996.

8.  US Patent 5867729, System for reconfiguring a keyboard configuration in response to an event status information related to a computer's location determined by using triangulation technique, Invented by John Mensick, Oct 1999

9. US Patent 5954437, "Computer keyboard system enabling users to locate keys of letters, radicals and phonetic symbols quickly", Wen-Hung, Sep 99

10. US Patent 6382854, Computer keyboard enhancement kit., Morelos, May 2002

11. US Patent and Trademark Office (USPTO) site, http://www.uspto.gov/